\documentclass[preprint,showpacs,preprintnumbers,amsmath,amssymb]{revtex4}
\usepackage[english]{babel}
\usepackage{bm} % bold math
\usepackage{graphicx}% Include figure files

\newcommand{\beq}{\begin{equation}}
\newcommand{\eeq}{\end{equation}}
\newcommand{\bea}{\begin{eqnarray}}
\newcommand{\eea}{\end{eqnarray}}

\begin{document}
\selectlanguage{english}

\title{Offsets in Electrostatically Determined Distances: Implications for Casimir Force Measurements}

\author{\firstname{S.~K.}~\surname{Lamoreaux}}
\email{steve.lamoreaux@yale.edu}
 \affiliation{Yale University, Department of Physics, P.O. Box 208120, New Haven, CT 06520-8120}
\author{\firstname{A.~O.}~\surname{Sushkov}}
\email{alex.sushkov@yale.edu}
 \affiliation{Yale University, Department of Physics, P.O. Box 208120, New Haven, CT 06520-8120}

\begin{abstract}
The imperfect termination of static electric fields at semiconducting surfaces has been long known in solid state and transistor physics.  We show that the imperfect shielding leads to an offset in the distance between two surfaces as determined by electrostatic force measurements.  The effect exists even in the case of good conductors (metals) albeit much reduced.
\end{abstract}

\maketitle

It has been long known that a static electric field applied external to a semiconductor will not be perfectly internally shielded.  We previously considered the case of Debye screening, which predicts a penetration depth of 0.7 $\mu$m for Ge, and 24 $\mu$m for Si (intrinsic undoped materials at 300 K) \cite{debye}, which leads to distance offset when the force due to a voltage applied between the plates is used for distance determination. The distance correction given by the twice the Debye length divided by the (bare) dielectric constant.  Our measurements with Ge \cite{kim} did not support such a large distance correction, and indeed it has been long known that surface states lead to a shielding of the internal field.  The first attempts to build a field effect transistor showed an internal field at least 100 times smaller than expected based on bulk material considerations.\cite{shockley}  Bardeen provided a complete explanation of the importance of surface states, and the basic premises are well established fundamental considerations in transistor physics.\cite{bardeen}  These effects largely explain why no distance correction was observed in our Ge experiment. The analysis presented here, which included band bending and surface state effects, shows that the simple Debye screening treatment of this problem is an oversimplification.

We are reconsidering the problem because in our recent attempts at measuring the Casimir force between high resistivity ($> 10 {\rm k}\Omega$cm) Si plates, we see a distance offset of between 60 nm and 600 nm, depending on how the plates were cleaned.  The sensitivity of the effect to surface condition suggests that the surface states are likely the most important part of the offset correction. In our study of the effect, it has become apparent that distance offsets are probably affecting all Casimir experiments to date, and is an effect that needs to be considered among the panoply of systematic effects that are only recently being acknowledged.  It is interesting to note that all of these systematic effects (roughness, vibration, surface patch potentials, and now the offset effect considered here) all lead to an apparent {\it increase} in the Casimir force compared to its true value at a given distance as determined electrostatically.  

The basic physical principle is shown and explained in Fig. 1.  It is usually assumed that the charge density in the depletion region is constant ($n_d$) over its width $d_1$.\cite{dekker} This approximation is very good in the case where there Fermi level is well below the conduction band; raising the potential of the conduction band a small amount essentially empties it.   

Shown in Fig. 2 are the related processes that are important for an electrostatic distance calibration.  We consider a case where the system is in equilibrium when a potential $V_0$ is applied between the plates, and then ask what happens when $V_0$ is changed by $\delta V$.  

First, when $\delta V=0$, we have (in units with $\epsilon_0=1$, and the electron charge $e=1$)
\begin{equation}
\sigma_0 d_0 + V_1=V_0
\end{equation}
by energy conservation and
\begin{equation}
\sigma_0+\sigma_1-n_d d_1=0
\end{equation}
by charge neutrality, 
where $\sigma_0$ is the charge on the perfectly conducting plate, $\sigma_1$ is the surface state charge on the semiconducting plate, $d_0$ is the physical plate separation.  Integrating the electric field from the surface into the bulk, using Poisson's equation, indicates that 
\begin{equation}
V_1=\alpha d_1^2
\end{equation}
where $\alpha\approx n_d/\kappa$ with $\kappa$ the bare dielectric constant.  Furthermore, we have 
\begin{equation}
\sigma_1= (E_F-V_1)n_s
\end{equation}
where $n_s$ is the surface density of states ({\it e.g.}, electrons/cm$^2$volt in appropriate units). We thus find that
\begin{equation}
\sigma_0=-(E_F-V_1)n_s+n_d\sqrt{V_1/\alpha}
\end{equation}
or
\begin{equation}
\sigma_0=-(E_F-(V_0-\sigma_0 d_0)n_s)+n_d\sqrt{(V_0-\sigma_0 d_0)/\alpha}
\end{equation}
which gives the relationship between $\sigma_0$ and $V_0$ which can be used to find the net effective {\it differential} capacitance.  It is easiest to change $V_0$ by $\delta V$ and determine $\delta V_1$, $\delta d_1$, $\delta \sigma_0$ and $\delta \sigma_1$, with fixed $n_d$, $E_F$, $n_s$, and $d_0$.  From charge neutrality, we have
\begin{equation}
\delta \sigma_0-n_s\delta V_1-n_d\delta d_1=0.
\end{equation}
Furthermore,
\begin{equation}
\delta d_1={1\over 2}{\delta V_1\over \sqrt{\alpha V_1}}
\end{equation}
and
\begin{equation}
d_0\delta\sigma_0+\delta V_1 =\delta V.
\end{equation}
We thus arrive at
\begin{equation}
\delta \sigma_0+n_s d_0\delta\sigma_0+d_0 n_d(4 \alpha V_1)^{-1/2}=n_s\delta V +n_d(4\alpha V_1)^{-1/2}\delta V
\end{equation}
which gives a differential capacitance (per unit area) of
\begin{equation}
C={\delta \sigma_0\over \delta V} ={n_s+n_d(4\alpha V_1)^{-1/2}\over 1/d_0 +n_s+n_d(4\alpha V_1)^{-1/2}}\times {1\over d_0}.
\end{equation}
Evidently, this equation can be written in a form
\begin{equation}
C={C_A C_B\over C_A + C_B}
\end{equation}
where 
\begin{equation}
C_A={1\over d_0}=C_{gap}
\end{equation}
and
\begin{equation}
C_B= n_s+n_d (4\alpha V_1)^{-1/2}=C_{surface}+ C_{bulk}=C_{maximum}={1\over d_{offset}}
\end{equation}
which is the maximum capacitance that can be observed as $d_0\rightarrow 0$. It should be noted that the offset distance does not depend on $d_0$; the capacitance between a spherical and flat surface has the same offset as that for flat plates, in the limit that the proximity force approximation applies.

Taking into account surface and bulk (space charge) effects shows that the net (differential) capacitance is lowered due to the parallel combination of $C_{surface}$ and $C_{bulk}$ in series with $C_{gap}$.  Thus, the plates are {\it closer} than the distance given by an electrostatic calibration. When both plates are made of the same semiconducting material, the distance offset calculated here is simply multiplied by 2. 

It is almost impossible to know the parameters required to calculate the surface and bulk differential capacitances.  For our recent measurements with Si, we have measured the distance offset directly in the experiment ({\it in situ}) by performing an electrostatic distance determination, then measuring how far the plates must be moved until they touch.  A more accurate determination of the offset was possible by firmly mounting the plates, with one on a translation stage.  The offset was determined by measuring the capacitance as a function of distance until the plates touched, which could be readily electrically determined.  A maximum capacitance, at zero plate separation as determined by the electrical contact, was directly measured, and found to be 62 $\pm$ 5 nm.  After further cleaning of the plates, an {\it in situ} measurement shows a 600 nm distance offset.  

We further note that even in the case of good conductors, there is a space charge (bulk) correction to the capacitance.  In \cite{wallmark} (Sec. 3-2.4) it is shown that in the case of a parabolic conduction band, the effective distance offset is about 0.1 nm per plate.  This means that for distances of approximately 100 nm, in a force gradient type experiment between a flat and smooth surface, noting that the electrostatic distance offset means the plates are closer than expected, the correction is
\begin{equation}
{\delta F\over F}=-4 {\delta d\over d}=-4\times (-0.2)/100\approx 1\%
\end{equation}
which is at the level of the claimed accuracy of several experiments.

Because the offset is very sensitive to the physical surface properties, it would appear as prudent to either directly measure the offset, or include it as an adjustable parameter in comparing theory to experiment. We have found that it is simple to measure the maximum capacitance that occurs just before the plates come into physical (electrical) contact.  This distance needs to be corrected by the surface roughness, and the measurements must be done at sufficiently low frequency for equilibrium to be established, however, the measurements are straightforward.

\begin{figure}
\includegraphics[width=6in]{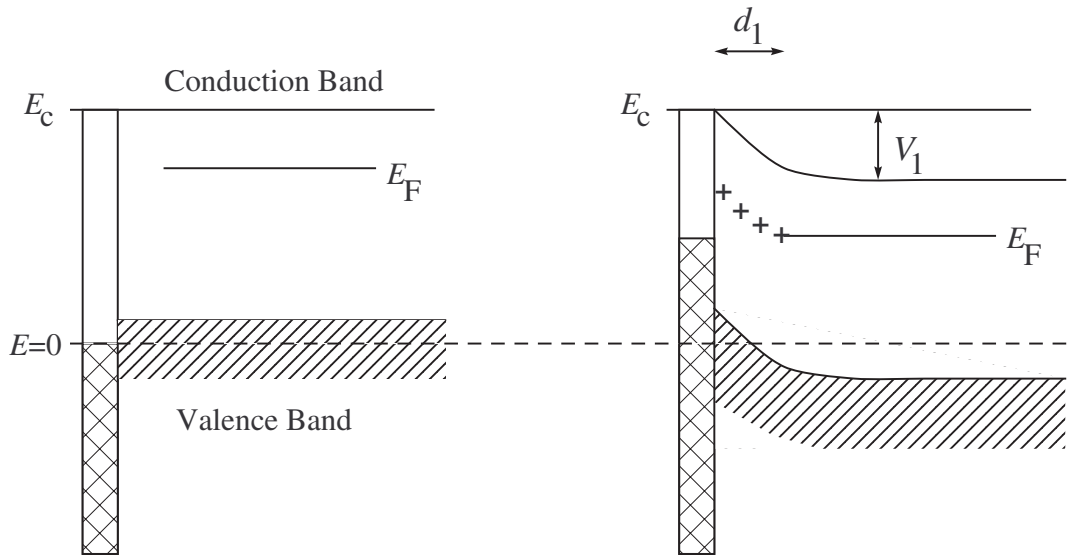}
\caption{On the left is shown a non-equilibrium distribution of electrons; taking $E=0$ as an ``uncharged" surface, the surface state energy does not equal the Fermi energy $E_F$ which lies between the valence and conduction bands for a semiconductor.  To establish equilibrium, electrons flow from the bulk to surface states until the surface states are filled to $E_F$, a, leaving a depletion, or ``space charge" region of depth $d_1$. This redistribution of charge causes the bands to bend, as shown.}
\end{figure}

\begin{figure}
\includegraphics[width=6in]{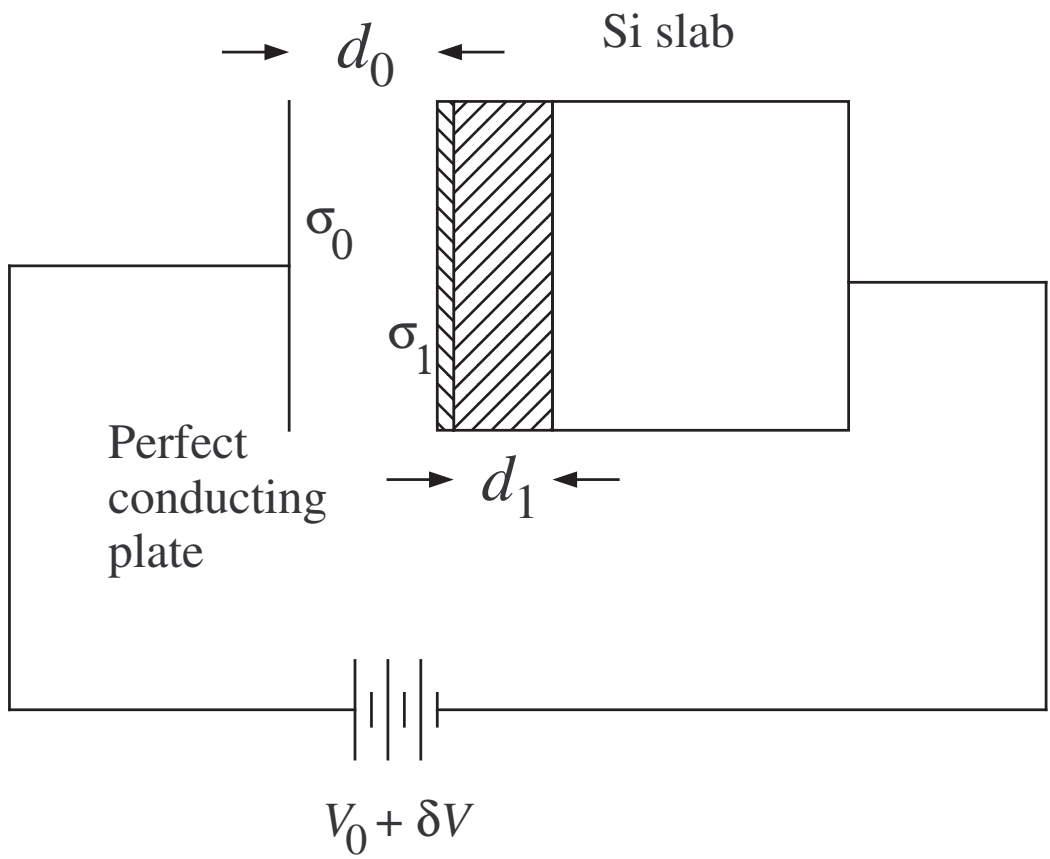}
\caption{Schematic model of electric field penetration through the  surface of a semiconductor.}
\end{figure}

\end{document}